\documentclass[prl,preprintnumbers,tightenlines,twocolumn]{revtex4-1}
\usepackage{braket}
\usepackage{array}
\usepackage{dsfont}
\usepackage{amsmath}
\usepackage{pifont}
\usepackage[normalem]{ulem}
\usepackage{amssymb}
\usepackage{xcolor}
\usepackage{amsfonts}
\usepackage{mathtools}
\usepackage{graphicx}
\usepackage{dcolumn}
\usepackage{bm}
\usepackage{multirow}
\usepackage{hhline}
\usepackage{siunitx}

\usepackage{leftidx}
\usepackage{color}
\usepackage{mathtools}
\usepackage{MnSymbol}
\usepackage[mathscr]{eucal}
\usepackage[german,english]{babel}
\usepackage[capitalize]{cleveref}
\def\jm#1{\textcolor{black}{#1}}
\def\nn {n}

\begin{document}
	\title{Rydberg states with a liquid core}
	\author{Juan Carlos Acosta Matos}
	\author{P. Giannakeas}
	\author{Jan M. Rost}
	\affiliation{Max-Planck-Institute for the Physics of Complex Systems, N\"othnitzer Str.\ 38, D-01187 Dresden, Germany }

	\begin{abstract}
		\noindent
		We develop a self-consistent approach that provides an explicit potential for a Rydberg electron whose ionic core consists of a polarizable medium, typically realized with  superfluid droplets. The electron's motion remains separable in spherical coordinates, but the radial force exerted by the droplet breaks degeneracy of the angular momentum states non-perturbatively. The ensuing electron spectrum reveals intriguing properties dependent on droplet size and electron excitation. Deviations of the polarizable medium from the continuous spherical distribution can be taken into account as a perturbation of this redefined Rydberg dynamics. We discuss specific but paradigmatic examples for superfluid helium and also propose a way to probe droplet properties including its possible crystallized fraction through stimulated transitions of the Rydberg electron. 
	\end{abstract}
	\maketitle
	
Since the empirical discovery of emission and absorption lines by Balmer and Rydberg the so-called Rydberg states have found continued interest. Important conceptual advances include the development of quantum defect theory for precision spectroscopy \cite{gefa+79,Ga94} in molecules and atoms. Advanced cooling techniques have given rise to ultracold gases, and in that vein Rydberg atoms have evolved into experimental platforms for quantum technologies and high precision spectroscopy thanks to sophisticated methods to control external fields \cite{scag+23}.  Ultracold environments also allow for the sustained creation of giant atoms with very large electronic Rydberg orbits of micrometer diameter (principal quantum number $n{=}100$ and higher). Such states can bind  a neutral atom to form an extremely long-range molecule \cite{grdi+00,bebu+09,hamilton2002shape,dressedgiannakeas2020,bebu+10} and even several atoms can become bound to the Rydberg atom \cite{luro17} eventually leading to a polaron picture of a Rydberg impurity embedded in a cold gas \cite{casc+18,duei24,chowdhury2024ion}, for recent reviews see \cite{eipe+16,duka+24}.
	
Beyond the traditional ionic or molecular core, large Rydberg orbitals can encompass a large finite system such as a droplet or a nano-crystallized structure \cite{hunterRydbergComposites2020},  which can constitute a strong distortion of the hydrogenic electron orbitals.  Yet, for large enough Rydberg orbits, the enclosed system will look spherical, no matter what its actual shape or structure is, and the spherical symmetry of the Rydberg electron density will be preserved. Even for the most anisotropic enclosed systems, their spherically averaged continuous density contains the main part of the non-perturbative distortion of the hydrogenic spectrum, while the deviation from this droplet density can be described perturbatively. Likewise, typical Rydberg orbitals are spread out enough in space to neglect their back-action on the density of the droplet. If this is not the case, the correct droplet density can be determined self-consistently by iteration.

To be specific, we consider a Rydberg atom where the core consists of an ion \jm{in the center of}  a compact environment which we assume to be confined within a sphere of radius $R_d$.  It interacts with the Rydberg electron through the Fermi pseudopotential~\cite{fermiNuovoCim1934}, characterized by the $s$-wave scattering length $a_s$. The environment may be anisotropic, either through ion-induced structuring manifested, e.g., for helium droplets in the formation of snowball and supersolid layers~\cite{matosPhys.Rev.Lett.2025,mateoSolvationIntrinsicPositive2014,mullerAlkaliheliumSnowballComplexes2009,chikinaPRB2007,zunzunegui-bruACS2023a,bartlJPCA2014,slavicekPCCP2010,tramontoPathIntegralMonte2015,galliPathIntegralMonte2011}, or when modeled as an ensemble of static point-like scatterers with a fixed regular or irregular spatial pattern~\cite{hunterRydbergComposites2020}. \jm{The charged impurity located off center constitutes a second, qualitatively different way to break the spherical symmetry, which we will address later}.  In a simple but crucial step we decompose the density $\rho(\mathbf{R})$ of the environment into an isotropic $\rho(R)$ and a $\tilde{\rho}(\mathbf{R})$ component, $\rho(\mathbf{R}) = \rho(R) + \tilde{\rho}(\mathbf{R})$.  Far away from the core, relevant for very large Rydberg orbits, only the isotropic component will survive.
	
 In superfluid helium droplets doped with alkali ions~\cite{matosPhys.Rev.Lett.2025}, the isotropic component can be identified with the pure droplet density $\rho(R)$ without impurity. In cases where the anisotropy extends throughout the droplet, e.g., if the atoms appear as point-like scatterers, the isotropic component can be taken as the spherical average of the density,  $\rho(R) = \frac{1}{4\pi} \int \rho(\mathbf{R})d\Omega $. In both scenarios, the anisotropic contribution $\tilde{\rho}(\mathbf{R})$ accounts for the fluctuations on top of $\rho(R)$.

With the decomposition of the density, also the Hamiltonian  $H = H_0 + h$ splits naturally into two parts. The anisotropic part
	\begin{equation}
		h =  2 \pi a_s  \int d^3\mathbf{R} \,\,\, \tilde\rho(\mathbf{R}) \ket{\mathbf{R}} \bra{\mathbf{R}} 
		\label{eq:H_pert}
	\end{equation}
	can be seen as a perturbation to the spherically symmetric part, 
	\begin{equation}
		H_0 = \textstyle{\frac12}\nabla^2 + U(r) + 2 \pi a_s \rho(r)\,,
		\label{eq:H0}
	\end{equation}
	which accounts for the effect of the ‘‘liquid" background (determined by droplet size and average density) on the electronic cloud. We use atomic units unless stated otherwise.
	In the present context the ion-electron interaction is given by the Coulomb potential $U(r)=- 1/r$.
	The isotropic contribution $2 \pi a_s \rho(r)$ in ${H}_0$ strongly modifies the Rydberg spectrum, representing the dominant effect of the droplet. Since the spherical symmetry is preserved, the eigenstates of the droplet Hamiltonian $H_0$ remain separable with a complete set of quantum numbers.  These
	\textit{droplet-dressed Rydberg states} (DDR  states)  $|N\ell m\rangle$   have the spatial representation 
\begin{equation}
    	\braket{\mathbf{r}|N \ell m} 
	= \frac{u_{N \ell}(r)}{r}\,Y_{\ell m}(\theta,\phi)\,,
\end{equation}
with spherical harmonics $Y_{\ell m}$. 
The radial wave function $u_{N\ell}(r)$ is determined by the effective radial potential $V^{\rm eff}_\ell(r)$ made up of the centrifugal barrier for a given angular momentum $\ell$, the Coulomb attraction and the droplet potential,
\begin{equation}
	\label{eq:Uell}
	V^{\rm eff}_\ell(r) = \frac{\ell(\ell+1)}{2r^2}-\frac1r +2 \pi a_s \rho(r)\equiv   U_\ell(r) +2 \pi a_s \rho(r) \,.
\end{equation}

We expand the radial DDR wavefunction in the radial hydrogen basis $\{\phi_{\nn\ell}(r)\}$, 
\begin{equation}
	u_{N \ell}(r) = \sum_{\nn} C^{\nn}_{N \ell}\, \phi_{\nn \ell}(r)\,.
	\label{eq:radial_ddrs}
\end{equation}
This representation is particularly convenient, since the coefficients $C^{\nn}_{N \ell}$ quantify directly the droplet-induced mixing of hydrogen states with principal quantum numbers $n$. The mixing can be large beyond any perturbative limit,  as can be seen in Fig.~\ref{fig:spectrum_rad250}. On the other hand, for large enough  $N$ the eigenstates will merge with the ones of hydrogen in the surviving potential $U_\ell(r)$, i.e., $C^\nn_{N \ell}\to \delta_{\nn N}$ for $N\to\infty$. For this reason, we will use the hydrogen energies $E_n$  and principal quantum number $n$ as a reference.

\begin{figure}[h]
	\centering
	\includegraphics[width=0.99\columnwidth]{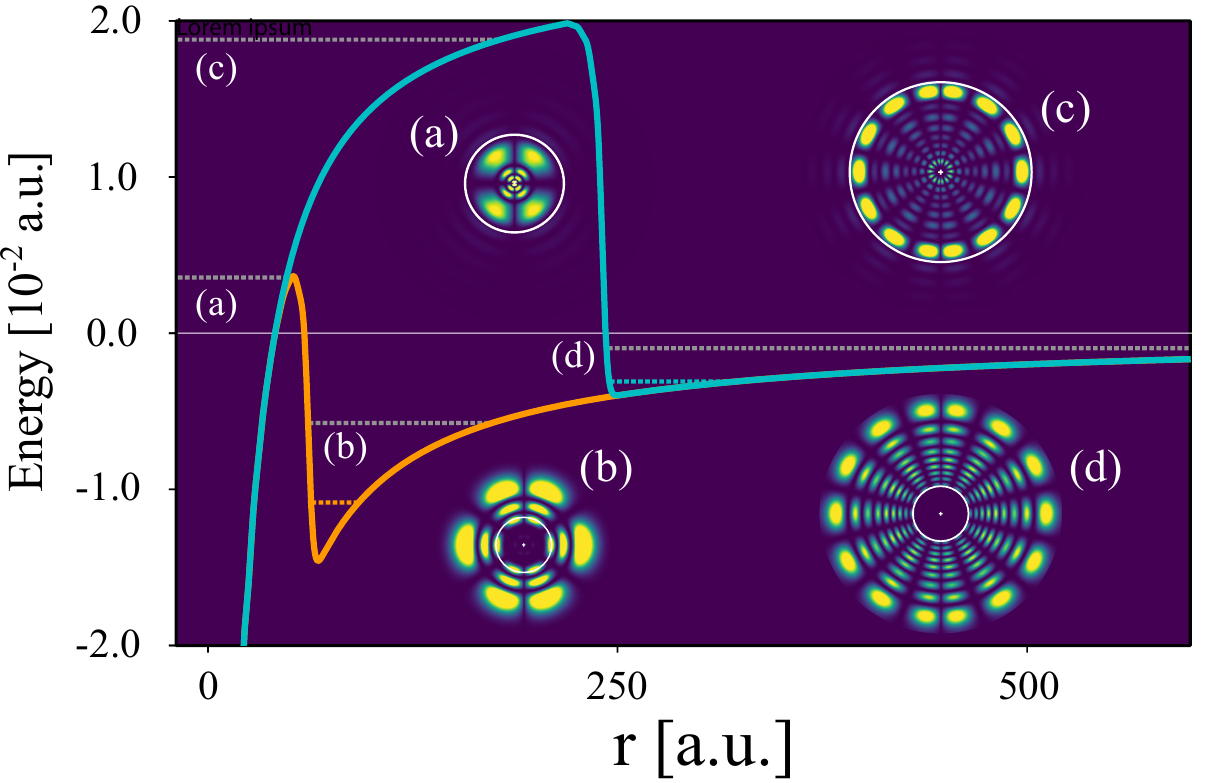} 
		\caption{Radial potential of \cref{eq:H0} for droplets of radius $70\,a.u.$ (orange) and $250\,\rm{a.u.}$ (cyan). Horizontal dashed lines (orange and cyan) indicate the oDDR ground state energy for each droplet. Insets show cylindrical-coordinate surface plots of the Rydberg electron probability for  the quasi-bound iDDR states (a,c)  and the bound oDDR states (b,d), where the white circle marks the droplet boundary.
        The  quantum numbers $\ket{\bar n,\ell,m} = \ket{5,1,0}$ and $\ket{12,6,0}$  for (a) and (c), and  
         $\ket{\nu,\ell,m} = \ket{5,1,0}$ and $\ket{15,6,0}$ for (b) and (d) reflect the   nodal patterns  of the respective states. The eigenenergies of the states are shown as dashed gray lines. Note that, the corresponding effective radial potentials have in addition to the ones shown for $\ell=0$ centrifugal barriers $\ell(\ell+1)/(2r^2)$.}
	\label{fig:potential}
\end{figure}

As a concrete illustration but without loss of generality, we present DDR states and spectra for a helium droplet doped with an alkali Rydberg atom, where the helium-electron interaction is repulsive, $a_s = 1.186\,\rm{a.u.}$~\cite{fedusAtoms2021}. 
Figure~\ref{fig:potential}  shows the radial  potential of \cref{eq:H0}, including both, the Coulomb attraction to the ionic core and the repulsive droplet barrier. 
In general, Rydberg systems with a liquid core feature two universal classes of DDR states. oDDR states are primarily localized {\it outside} the droplet ((b) and (d) in Fig.~1) and have always negative energies (true bound states)  although with a quite intricate structure (see Fig.~2b) which  we will  discuss shortly.  \jm{They are akin to   states of an ``excimer atom" proposed  in a pioneering  analytical study decades ago \cite{golov1993new}}. On the other hand iDDR states are primarily localized {\it inside} the droplet. To distinguish both classes we designate the radial quantum number ($N$) for iDDR states with $\bar n$ and  for oDDR states with $\nu$ \footnote{Formally, all bound states with a given $\ell,m$ could be enumerated according to their number of radial nodes. The classification with $\nn$ and $\nu$, however, is more physical, as it reflects the number of nodes in a spatial region where the eigenstate has significant amplitude.}.
In Fig.~\ref{fig:potential} the lowest oDDR energy  is marked with dashed  orange and cyan lines for the small and larger droplet, respectively.  

The electron density of iDDR states can  even be confined to a region
where the droplet has a constant angular averaged density  $\bar \rho$ (see Fig.~1(a,c)). Consequently, those states have upwardly shifted hydrogen eigenenergies 
\begin{equation}
	e_\nn = E_\nn + 2\pi a_s \bar{\rho}\equiv -\frac{1}{2 \nn^2} + 2\pi a_s \bar{\rho}\,.
	\label{eq:quasi-bound_energies}
\end{equation}
For helium with  $\bar{\rho} \approx 3.3 \times 10^{-3} \rm{a.u.}$ this means $e_\nn<0$ for $\nn \leq 4$, while iDDR energies  are shifted into the quasi-bound region with $e_\nn>0$ for $\nn \geq 5$ (Fig.~2b). Depending on the droplet size there is a critical excitation $\bar n$  indicative of a transitional regime \footnote{For the droplet with $R_d=250 \, \rm{a.u.}$, the critical manifolds  are $11$ and $12$.}, where the Rydberg wavefunction 
neither fully fits into the droplet nor is completely outside the droplet.  While the typical 
iDDR states are almost pure hydrogen states with a  dominant contribution from a single hydrogenic $\bar n$ (for a droplet of radius $250 \, \rm{a.u.}$, this happens for $\bar{n} \leq 10$),
 transitional states ($\bar{n} \geq 11$) are spread over several manifolds $n$ and have typically shorter lifetimes. Hence, their energies deviate from \cref{eq:quasi-bound_energies}, see Fig.~\ref{fig:spectrum_rad250}(a). 
 
\begin{figure}[h]
	\centering
	\includegraphics[width=0.99\columnwidth]{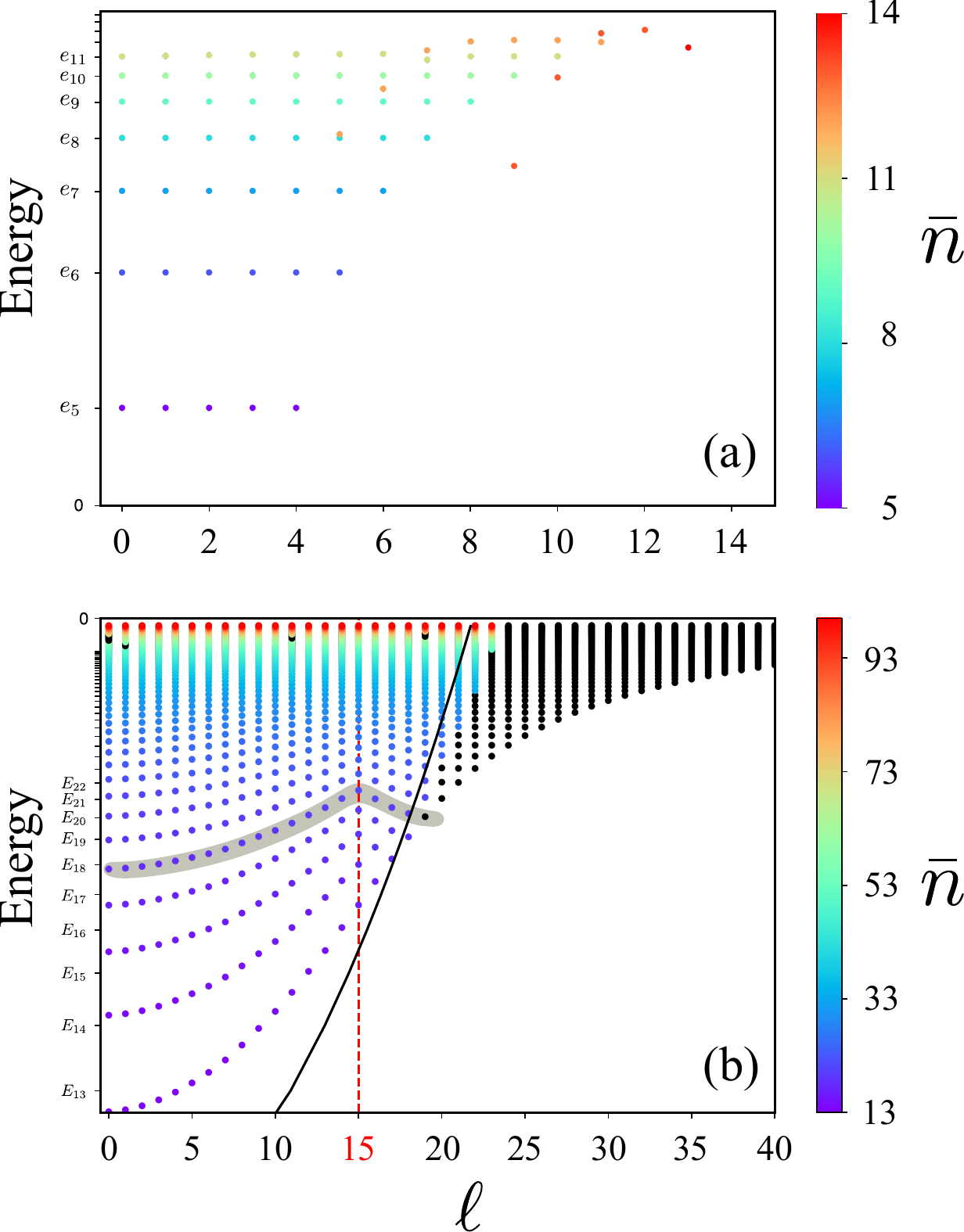} 
	\caption{DDR spectrum for a droplet of radius $250\,\rm{a.u.}$ for the quasi-bound states (a) and the bound states (b). Ticks on the vertical axis are related by  \cref{eq:quasi-bound_energies}. Colors indicate the dominant hydrogenic manifold, $\bar{n}$, contributing to each state. Black points denote states with a single 
		hydrogenic component $C^\nn>0.99$.  Their onset  is estimated from  \cref{eq:estimator_lmin_for_hydrog_states} (black line). $\ell=15$ marks the estimated kink position in the spectrum (see  \cref{eq:estimator_l_kink}).  }
	\label{fig:spectrum_rad250}
\end{figure}

We proceed to discuss in more detail the oDDR states.  For the $R_d = 250\,\rm{a.u.}$ droplet of  Fig.~\ref{fig:spectrum_rad250}(b), they begin with $\bar n = 13$. Recall that oDDR states are bound states fully localized in the effective radial potential wells outside the droplet.
The energies labeled black in Fig.~\ref{fig:spectrum_rad250}(b) are purely hydrogenic. To estimate the maximum angular momentum $\ell$ for which the droplet still distorts hydrogen states as a function of $R_d$, we determine the classical turning point
\begin{equation}
	\ell(\nn,R_d) = \frac{\sqrt{\,1 - 4(R_d^2/\nn^2 - 2R_d)}-1}{2}
	\label{eq:estimator_lmin_for_hydrog_states}
\end{equation}
of the hydrogenic radial wavefunction at the droplet surface $U_\ell(R_d) = - \frac{1}{2\nn^2}$  with $U_\ell$ from \cref{eq:Uell}.
For a given droplet size, the maximum $\ell$ is obtained in the limit $\nn \to \infty$,  
\begin{equation}
	\ell_M(R_d) = \frac{\sqrt{\,1 + 8R_d}-1 }{2}.
	\label{eq:estimator_l_Max}
\end{equation}
 Shown as the black line in Fig.~\ref{fig:spectrum_rad250}(b), this simple estimator predicts approximately the onset of hydrogenic states for which the droplet is effectively transparent. $\ell_M$ provides a useful criterion to drastically reduce the number of basis states in \cref{eq:radial_ddrs}, which can be restricted to  maximum $\ell$ slightly above $\ell_M(R_d)$. 

Fig.~\ref{fig:spectrum_rad250}(b)  reveals that the spectrum is organized in branches with a characteristic shape.  The branch highlighted by  gray-shades contains $20$ states, distinguished by different angular momenta   $0\le \ell\le 19$ ending in the purely hydrogenic state $\phi_{\nn=20,\,\ell=19}(r)$. It gives rise to the branch label $\bar{n} = 20$, which provides a  systematic link to the iDDR states  and simultaneously indicates the dominant hydrogenic manifold contributing to states in the branch, see color codes in Fig.~\ref{fig:spectrum_rad250}. Branches at higher energies contain more states. Along a branch radial nodes get exchanged for angular nodes with increasing angular momentum $\ell$.   For pure hydrogen without the droplet, these states would appear as $20$   points on the horizontal line $E_{20}$, reflecting full $\ell$-degeneracy. In the presence of the droplet, however, strong $\nn$-mixing creates the branches with a characteristic kink (the maximum energy of the branch) at $\ell = \ell_d$. 
For $\ell \leq \ell_d$, the radial wave function is subject to the combined influence of the droplet induced potential and the centrifugal barrier, see \cref{eq:Uell}. For growing $\ell$  
the centrifugal barrier begins to dominate and $U_\ell(r)$ alone develops a minimum at the droplet surface $R_d$. According to $\frac{dU_\ell(r)}{dr}|_{r=R_d} = 0$, this happens for
\begin{equation}
	\ell_d = \frac{-1 + \sqrt{\,1+4R_d}}{2}.
	\label{eq:estimator_l_kink}
\end{equation}
Note that this condition depends only on the droplet radius $R_d$, and therefore the position of the kink in the branches at $\ell_d$ is independent of the energy, respectively, number of states in the branch.
For $\ell > \ell_d$, the influence of the droplet on the states progressively vanishes, and their energies decrease with increasing $\ell$ until the pure hydrogenic spectrum is recovered.

We note in passing that in systems with an attractive droplet-electron potential, the spectrum should also exhibit a characteristic feature at $\ell_d$, marking the maximum influence of the droplet on the electron before the centrifugal barrier dominates and the fully degenerate hydrogenic spectrum is eventually recovered for large $\ell$. Consequently, both Eqs.~\eqref{eq:estimator_l_Max} and \eqref{eq:estimator_l_kink}, can be regarded as universal estimators, providing  physical insight and computational efficiency in the description of Rydberg excitations in systems with a liquid core. We conclude the discussion of the spectrum Fig.~2(b) by noting that increasing energies for a given $\ell$ are labeled by $\nu =1,...,\infty$ and represent eigenstates with increasing radial excitation in a single radial potential that contains the centrifugal barrier defined by $\ell$.

\begin{figure}[h]
	\centering
	\includegraphics[width=0.99\columnwidth]{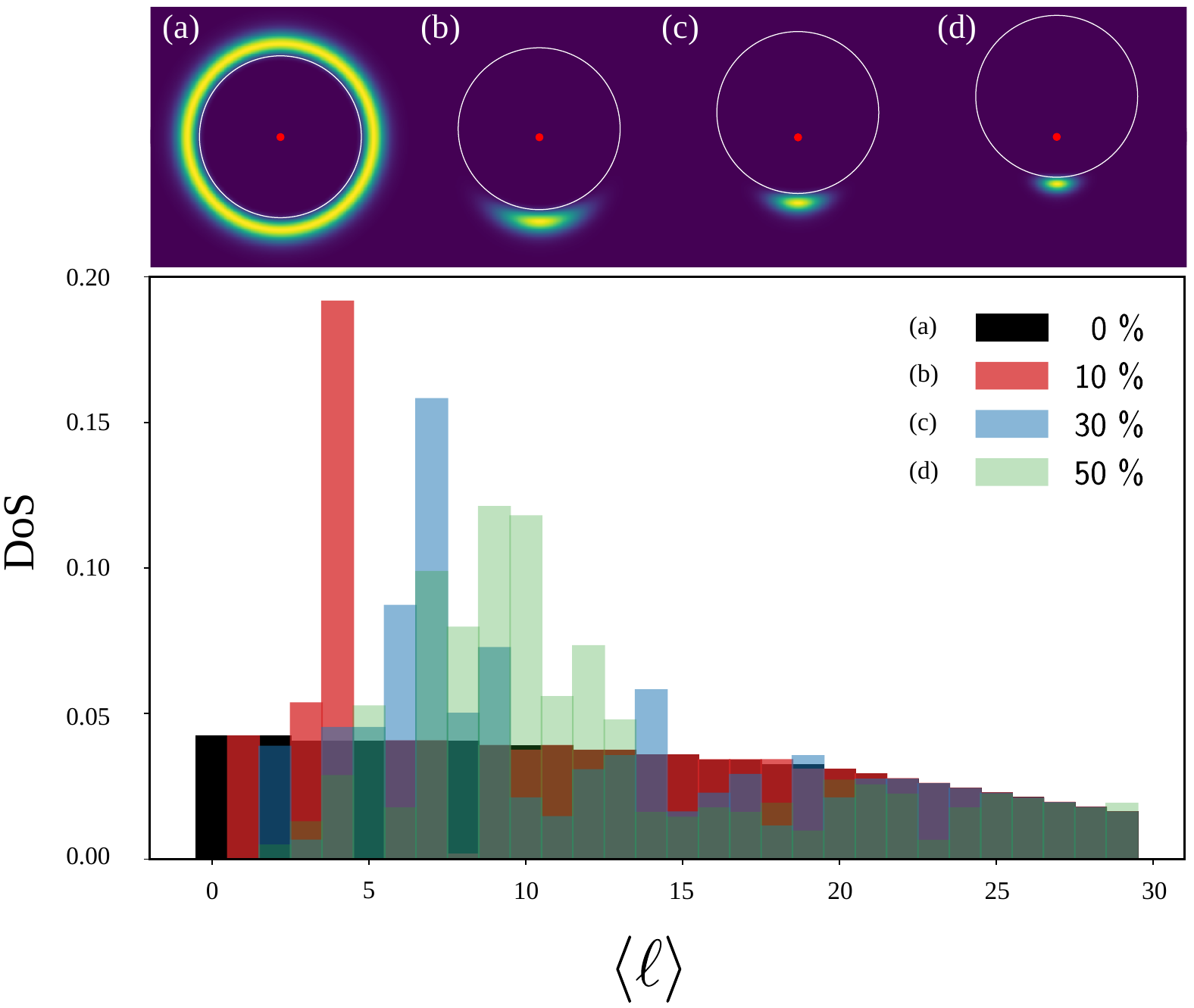} 
    	\caption{\jm{Spatial slice of the  electron density (a-d) for the lowest oDDR state  with increasing ion offset of 0 to 50\% from the center,    where 100\% correspond to an ion at the surface of the droplet with radius $R_d = 250\,\rm{a.u.}$. The histogram (area normalized to unity) shows the average angular momentum $\langle\ell\rangle$-density of $m{=}0$ states (DoS)  with respect to the ion center.} }
	
	\label{fig:ion_offset_DoS}
\end{figure}

oDDR states localize the electronic wave function outside the droplet region and therefore have no effect on the droplet density profile. In contrast, the electronic cloud of iDDR states resides inside the droplet and can act back on this profile.
The effect can be estimated from the ratio of the droplet-iDDR state interaction energy
to the energy of the droplet in the Rydberg volume. This ratio is a function of the droplet density, which in general must be determined self-consistently as it depends on the Rydberg state. The first iteration with a constant density $\bar\rho$, however, quantifies the relevance of the back-action. It can be easily determined from the constant shift in the eigenenergy of \cref{eq:quasi-bound_energies} that represents the iDDR-droplet interaction energy, relative to the total energy of the droplet in the Rydberg volume. The latter is roughly $v_n = 4\pi/3 r_n^3$  with $r_n \approx 3n^2$.  Hence, the relevance of the back-action of the electron on the droplet decreases rapidly with increasing $n$ according to the ratio 
\begin{equation}
	\delta(\nn) = \frac{1}{n^6}\left| \frac{a_s\bar\rho}{18 \, W[\bar{\rho}] }\right|\,,
	\label{eq:delta_n}
\end{equation}
where $\int_{v_n} W[\rho(\mathbf r)]d\mathbf r$ is an energy density functional of the droplet density.
A quantitative analysis for helium droplets is provided in the End Matter.

As laid out in the beginning, the spherical droplet with its continuous density together with \cref{eq:H_pert} can be taken as the basis to describe systems with anisotropic \jm{and spherical symmetry breaking} cores. 
\jm{An off-center ion is an obvious example. Reducing the symmetry from spherical to cylindrical it  leads to strong $\ell$-mixing in low-energy oDDR states. In contrast, most iDDR wave functions remain fully confined within the droplet rendering them insensitive to the ion displacement.  The evolution of the $\ell-$mixing for increasing displacement of the ion from the center  is illustrated in Fig.~\ref{fig:ion_offset_DoS}. One sees how the electron density of the lowest energy-oDDR state gets progressively localized angularly  (panels (a-d)). 
Analyzing the average angular momentum $\langle\ell\rangle$-density of lower lying oDDR states, the histogram reveals
how the peak density shifts to higher $\langle\ell\rangle$ and to a broader distribution with increasing displacement
of the ion from the center.   Since  symmetry broken states lift  the degeneracy in $\ell$, they are conveniently amendable to perturbation theory to include, e.g., spin-orbit effects for a specific impurity.  We provide a movie  \cite{suplemental_material} illustrating how the organziation of angular momentum progresses for increasing ion displacement, starting    with the ion in the center (Fig.~\ref{fig:spectrum_rad250}b).}

\jm{Note,  that experiments   Rydberg-excite an atom mostly close to the surface of the droplet \cite{von2005electronically,loginovPhys.Rev.Lett.2011,lacknerPhys.Chem.Chem.Phys.2011,loginovJ.Phys.Chem.A2011,pototschnig2017rydberg}. Thereby, they put the droplet into an extreme non-equilibrium situation which differs from our scenario of an equilibrated droplet density, no matter where the impurity sits.} 

As a second example of anisotropy we  consider a helium snowball, where the crystallized layers of a helium droplet with an ion at its center and the adjacent supersolid region provide the anisotropy \cite{matosPhys.Rev.Lett.2025}.  We know that large droplets strongly suppress Rydberg electron density inside the droplet for oDDR states. On the other hand, iDDR states can probe the interior of a  droplet and consequently also its anisotropy which extends roughly up to a radius of  $R_0 = 30 \, \rm{a.u.}$ for helium snowballs induced by alkali ions. Since we now need to diagonalize the anisotropic Hamiltonian in the full DDR basis, the estimator in \cref{eq:estimator_l_Max} is very helpful, as it tells us that we only need to include
states up to $\ell_M(R_0) \approx 7$ for the region $R\le R_0$. To be on the safe side we restrict the DDR basis to $\ell \leq 10$.  Eigenstates  $\lvert \alpha \rangle 
\;=\; \sum_{N\ell m} A_{N \ell m}^{(\alpha)} \,\lvert N \ell m \rangle$  are sorted in ascending order of energy and labeled by the index $\alpha$.
For these states we define the relative energy shift induced by the perturbation 
\begin{equation}\label{eq:shift}
	\mathcal{E}_{\alpha} 
	\;=\; \left\lvert\,1 \;-\;E_{\alpha}/E_{N \ell} \right\rvert,
\end{equation}
where the DDR state with the main contribution to the perturbed state $\ket{\alpha}$ provides the
reference energy $E_{N \ell}$. 
\begin{figure}[h]
	\centering
	\includegraphics[width=0.99\columnwidth]{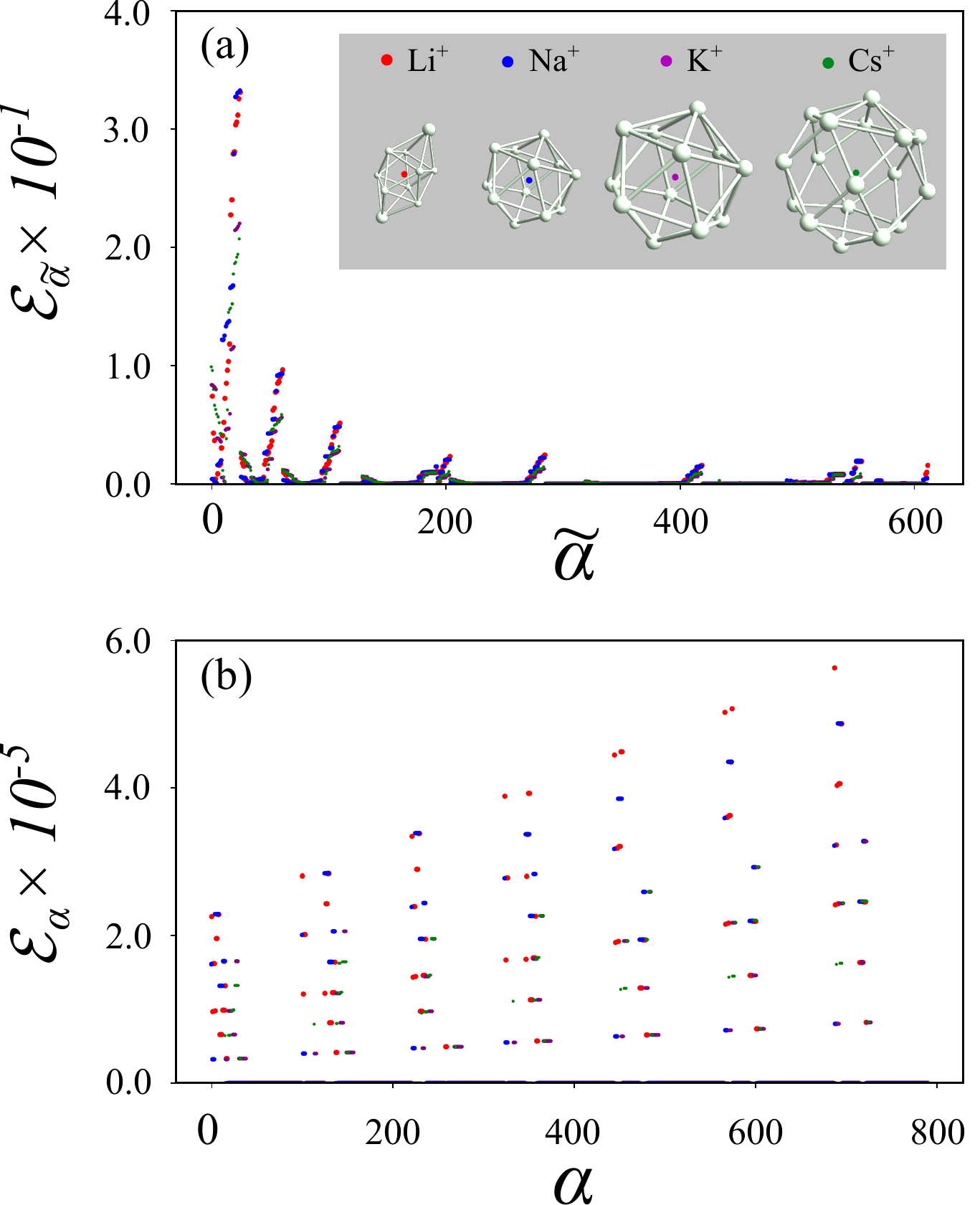} 
	
	\caption{Relative energy shift \cref{eq:shift} for a helium droplet of radius $R_d=250 \, \rm{a.u.}$ doped with different alkali ion impurities. (a) Quasi-bound states are labeled $\tilde{\alpha}$, and (b) bound states are labeled $\alpha$. For this illustration  atomic quantum defects are omitted as well as DDR basis states  with $\ell > 10$ (see text). \jm{The inset shows the  crystallized helium atoms of the first layer surrounding the ionic core for selected alkali ions. This first snowball layer exhibits the same icosahedral structure  for Na$^+$ and K$^+$, although the snowball of  K$^+$  has a larger radius.} }
	
	\label{fig:snowball_fingerprint}
\end{figure}

The anisotropic structures (snowball and supersolid layers) that form around ionic cores strongly depend on the ion species. For the alkali cores considered here, the perturbation induces relative energy shifts of up to $\sim 35\%$ for quasi-bound states and $\sim 0.006\%$ for bound states, see Fig.~\ref{fig:snowball_fingerprint}. States with low angular momentum are particularly sensitive to the details of the anisotropic helium structure, leading to species-dependent energy differences upon mixing. 
This opens the possibilty to experimentally probe the anisotropic droplet structure via the electronic  iDDR spectrum. Since its states are sufficiently long lived (see Fig. \ref{fig:end_matter_lifetimes} in End Matter) the spectrum can be accessed spectroscopically
by first photo exciting a bound oDDR electron to a highly excited iDDR state (sufficient dipole coupling is available, see Fig. S3 in End Matter), 
and subsequently stimulate transitions within the iDDR spectrum. \jm{Imperative is that after creation  of the ion, the droplet has time to adjust and equilibrate, through processes such as movement of the ion towards the center, forming of snowballs and evaporation of helium atoms.  In an experiment, this time could be gained by first ionizing the atom and later attaching a Rydberg electron through photo association, guaranteeing that the Rydberg electron represents a weak distortion.  }

In summary, 
we have presented a quantitative theory for Rydberg electrons with a liquid core \jm{realized} by, e.g., superfluid droplets, Bose-Einstein condensates   or supersolid media, independent of the specific anisotropy of the system. In all such cases, the construction of a spherically symmetric reference environment enables the definition of DDR (droplet-dressed Rydberg) states. They come in two classes, inner ones (iDDR) and outer ones (oDDR)  and capture the main distortions of the electronic cloud. The estimators introduced  in terms of critical angular momenta, namely $\ell_d$ for the maximal environmental influence, and $\ell_M$, indicating the onset of purely hydrogenic states, highlight universal spectral features. 

Our theory describes the interaction of Rydberg electrons with finite media in a simple fashion. If the medium represents a weak distortion, the oDDR energies can be described by quantum defects with the important difference, that oDDR states are eigenstates of a well-defined Hamiltonian specifying the liquid core and as such square integrable wavefunctions. Therefore, many applications of DDR states are possible including droplet-dressed Rydberg molecules, where a Rydberg atom with a liquid helium droplet core binds a neutral atom outside the droplet \cite{DropletMolecules}. In this case, the Rydberg electron, initially occupying a DDR state, mediates the binding and gives rise to ``dressed'' molecules whose properties reflect both, the highly excited electronic state and the underlying droplet environment.

\begin{acknowledgments}
\end{acknowledgments}

\bibliography{droplet}
\makeatletter 
\renewcommand{\thefigure}{S\@arabic\c@figure}
\makeatother

\makeatletter 
\renewcommand{\theequation}{S\@arabic\c@equation}
\makeatother

\setcounter{equation}{0}
\setcounter{figure}{0}

\section*{End matter}

\subsection{A. Semi-analytical estimate  of oDDR ground state energies}

\begin{figure}[h]
	\centering
	\includegraphics[width=0.9\columnwidth]{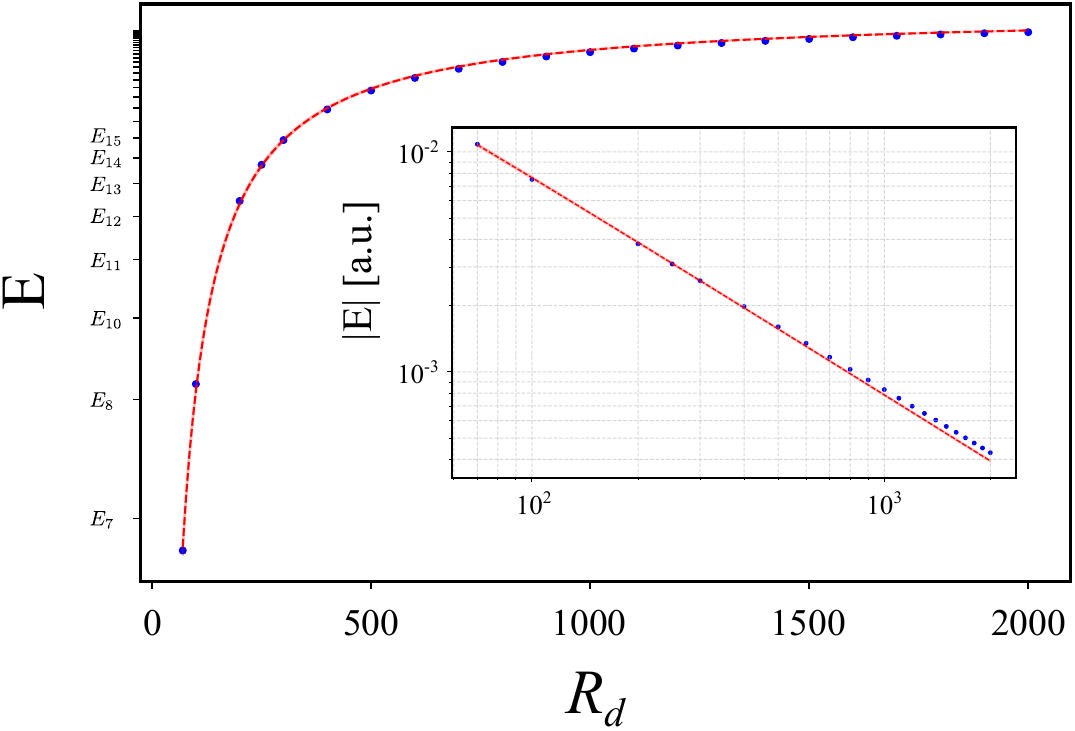} 
	\caption{Energy of the oDDR ground state as a function of droplet radius. Vertical ticks mark the unperturbed hydrogenic energy levels $E_\nn$. The red dashed curve shows the analytical estimate from \cref{eq:quantum_prediction_electron_energy}, while the blue points denote the computed oDDR ground state energies. The inset displays a log--log plot of the absolute oDDR ground state energy on the vertical axis.}

	\label{fig:ground_state_vs_drop_rad}
\end{figure}

With increasing 
droplet size, the energy of the oDDR ground state  $\ket{\nu=1,\ell=0,m=0}$ increases. This  oDDRg state is characterized by radial wave function near the droplet surface without nodes outside the droplet. From a classical perspective, the energy of a point-like electron orbiting at the droplet surface is expected to follow 
\begin{equation}
	E(R_d) = -\frac{A_c}{R_d},
	\label{eq:classical_electron_energy}
\end{equation}
with $A_c = 0.5\,{\rm a.u.}$. We parameterize the the oDDRg  energy as a function of droplet size with
\begin{equation}
	E(R_d) = -\,\frac{A}{R_d + a},
	\label{eq:quantum_prediction_electron_energy}
\end{equation}
where the parameters $A$ and $a$ account for quantum delocalization and the effective shift in the average electron position away from the droplet surface, respectively. Fitting this expression to the data [Fig.~\ref{fig:ground_state_vs_drop_rad}] for droplets with radii in the range $70–2000\,{\rm a.u.}$ gives
\begin{align}
	A &= (0.788120 \pm 0.007841)\,{\rm a.u.} \\
	a &= (3.164076 \pm 0.885842)\,{\rm a.u.}\,.
\end{align}
The resulting energy of \cref{eq:classical_electron_energy} is
in good agreement with the exact values, as can be seen in  Fig.~\ref{fig:ground_state_vs_drop_rad}. 

\subsection{B. Relative Electron-Droplet Energy } \label{sec:electron_droplet}

To quantitatively estimate the influence of the Rydberg electron's back-action  on the helium droplet as formulated in \cref{eq:delta_n}, we use the well-established Skyrme functional~\cite{dupont-rocInhomogeneousLiquid4HeDensity1990} to compute for a constant density $\bar \rho$:
\begin{equation}
\int_{v_n}	W[\rho(\mathbf r)] d\mathbf r = v_n \frac{\bar{\rho}^2}{2}(b + c \bar{\rho}^{\gamma}),
	\label{eq:end_matter_droplet}
\end{equation}
where $b = -1.89945 \times 10^{-2}$, $c = 4.68716 \times 10^4$,  $\gamma = 2.8$ and $v_n$ is given in the main text. Then, \cref{eq:delta_n} takes the explicit   
\begin{equation}
	\delta(\nn) = \frac{1}{\nn^6}\left| \frac{a_s}{9\,\bar{\rho} \bigl(b + c \bar{\rho}^\gamma \bigr)} \right|.
	\label{eq:end_matter_delta}
\end{equation}
As expected from \cref{eq:delta_n}, the back-action strongly decreases with
increasing Rydberg excitation   $\propto\nn^{-6}$. Importantly, it remains  small ($\delta \lesssim  10^{-1}$) for $\nn \geq 5$ and becomes negligible ($\delta \lesssim 10^{-3}$)  for $\nn \geq 10$, the typical range in Rydberg physics, therefore back-action effects were neglected in this work.

\subsection{C. iDDR lifetimes} \label{sec:lifetimes}

To estimate the lifetimes of the iDDR states, we employ the semiclassical approach of Ref.~\cite{gurvitzDecayWidthShift1987}. For quasi-bound states, the classical turning points inside the droplet, $r_0$ and $r_1$, and the classically forbidden region $r \in (r_1,r_2)$ are identified from the effective potential $V^{\rm eff}_\ell(r)$ in Eq.~\ref{eq:Uell}, with $r_2$ coinciding with the droplet radius.

For a given energy $E$, the local momentum is defined as $p(r) = \sqrt{2\bigl(E - V^{\rm eff}_\ell(r)\bigr)}$, which allows us to compute the action in the classically forbidden region,
\begin{equation}
    S = \int_{r_1}^{r_2} |p(r)|\,dr,
\end{equation}
together with the quasiclassical normalization 
\begin{equation}
    N = \int_{r_0}^{r_1} dr \, \frac{1}{p(r)} 
\cos^2\!\left(\int_{r_0}^{r} dr' \, p(r') - \frac{\pi}{4}\right).
\end{equation}

The estimated lifetime is then given by
\begin{equation}
    \tau = 4\,N e^{2S}.
\end{equation}

\begin{figure}[h]
	\centering
    
	\includegraphics[width=0.9\columnwidth]{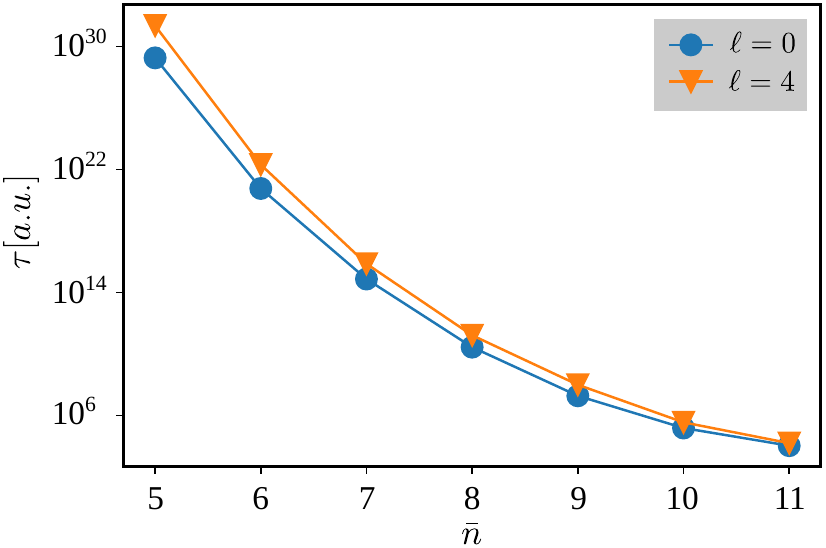} 
	\caption{Estimated lifetimes of selected iDDR states $\ket{\bar{n}\ell m}$ for $\ell=0$ and $\ell=4$, for a droplet of radius $R_d=250 ~ \rm{a.u.}$. As $\bar{n}$ increases, the states shift to higher energy and their electronic probability extends further inside the droplet towards the droplet boundary (see Fig.~\ref{fig:spectrum_rad250}(a) and Fig.~\ref{fig:potential} (c)). For $\bar{n}=11$ the lifetime is of the order of $0.1$ ps. }

	\label{fig:end_matter_lifetimes}
\end{figure}

iDDR states with low $\bar{n}$ typically exhibit longer lifetimes, which decrease rapidly as $\bar{n}$ increases, as shown in Fig.~\ref{fig:end_matter_lifetimes}. The spatial extent of the electronic state inside the droplet can be estimated as $\sim 3\bar{n}^2$; as $\bar{n}$ increases, the electronic wave function probes more deeply into the classically forbidden region, leading to enhanced tunneling and shorter lifetimes.

\subsection{D. oDDR to iDDR transition} \label{sec:transition}

The symmetry inherent in the approach described above guarantees the standard selection rules for transitions between DDR states, including those involving states with the electronic cloud localized either inside or outside the droplet. In particular, an initial state $\ket{i} = \ket{\nu \ell m}$ can couple to a final state $\ket{f} = \ket{\bar{n}\ell' m'}$ provided that $m = m'$ and $\ell' = \ell \pm 1$.

\begin{figure}[h]
	\centering
	\includegraphics[width=0.9\columnwidth]{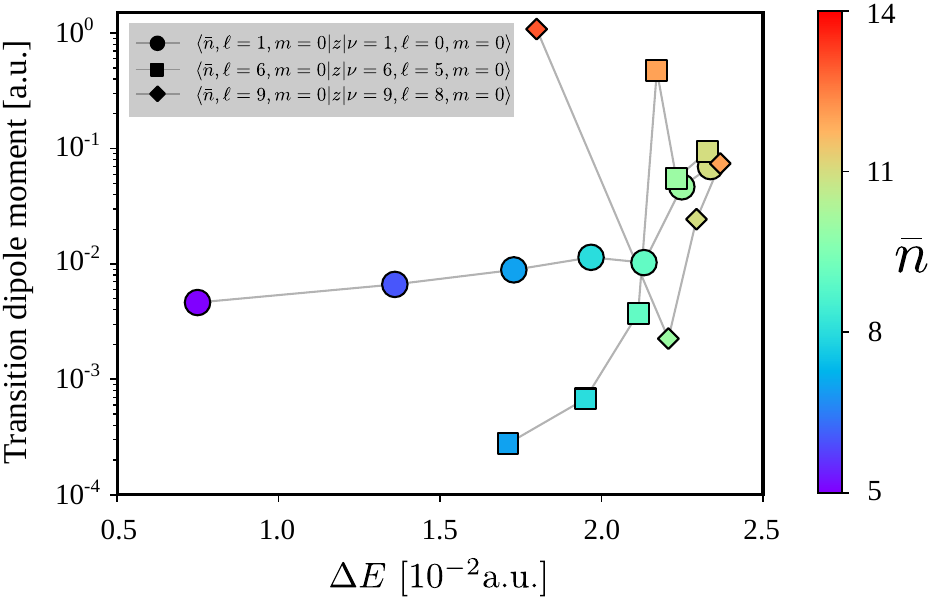} 
	\caption{Transition dipole moment $\braket{f|z|i}$ between an initial oDDR state (circular, square, and diamond symbols) and a target iDDR state, for a droplet of radius $R_d=250 ~ \rm{a.u.}$. The quantum numbers $\ell$ and $m$ are fixed by the selection rules, while the principal quantum number $\bar{n}$ is indicated by the color palette. The energy-difference axis corresponds to transition frequencies in the range $\sim 33$--$164$~THz.}

	\label{fig:end_matter_dipole_moment}
\end{figure}

However, enforcing transitions between states whose electronic probability is predominantly localized outside the droplet and those localized inside it introduces additional constraints. Direct optical coupling is efficient only to iDDR states with significant electronic probability density near the droplet boundary, whereas deeply localized iDDR states remain inaccessible from bound oDDR states through a single transition (see Fig \ref{fig:end_matter_dipole_moment} and Fig \ref{fig:spectrum_rad250}). Access to such deeply localized iDDR states may instead be achieved via a two-step process: first coupling an oDDR to an intermediate iDDR state with appreciable weight near the droplet surface, followed by a second transition to the target iDDR state.

\end{document}